# Point Group Symmetry Analysis of the Electronic Structure of Bare and Protected Metal Nanocrystals

Sami Kaappa,[1] Sami Malola[1] and Hannu Häkkinen[1,2,*]

[1] Department of Physics, Nanoscience Center, University of Jyväskylä, FI-40014, Jyväskylä, Finland
[2] Department of Chemistry, Nanoscience Center, University of Jyväskylä, FI-400014, Jyväskylä, Finland

**Abstract** The electronic structures of a variety of experimentally identified gold and silver nanoclusters from 20 to 246 atoms, either unprotected or protected by several types of ligands, are characterized by using point group specific symmetry analysis. The delocalized electron states around the HOMO-LUMO energy gap, originating from the metal s-electrons in the cluster core, show symmetry characteristics according to the point group that describes best the atomic arrangement of the core. This indicates strong effects of the lattice structure and overall shape of the metal core to the electronic structure, which cannot be captured by the conventional analysis based on identification of spherical angular momentum shells in the "superatom" model. The symmetry analysis discussed in this paper is free from any restrictions regarding shape or structure of the metal core, and is shown to be superior to the conventional spherical harmonics analysis for any symmetry that is lower than $I_h$. As an immediate application, we also demonstrate that it is possible to reach considerable savings in computational time by using the symmetry information inside a conventional linear-response calculation for the optical absorption spectrum of the $Ag_{55}$ cluster anion, without any loss in accuracy of the computed spectrum. Our work demonstrates an efficient way to analyze the electronic structure of non-spherical, but atomically ordered nanocrystals and ligand-protected clusters with nanocrystal metal cores and it can be viewed as the generalization of the superatom model demonstrated for spherical shapes ten years ago (Walter et al., PNAS **2008**, *105*, 9157).

**Keywords:** gold, silver, point group, symmetry, optical transitions

Symmetry lays the foundation to understand the electronic structure and spectroscopic transitions of small molecules, giving point group assignments of single-electron orbitals and dictating rules for allowed and forbidden dipole transitions between the quantum states.[1] Likewise, it can be used as an asset to predict properties of larger assemblies in the nanoscale even without explicit numerical computations, such as the high electronegativity of fullerene $C_{60}$ or metal/semiconducting characteristics of carbon nanotubes.

During the last decade, synthesis, structural determination and characterization of atom–precise ligand-protected metal nanoclusters have taken great leaps forward and currently over 100 structures of up to almost 400 metal atoms have been resolved (for recent reviews on experiments and theory, see Refs. 2-5). The current database of resolved structures reveals a multitude of shapes and atomic ordering in the metal cores, such as highly symmetric icosahedral or decahedral structures,[6-12] fcc-like packings,[13-15] and strongly non-spherical shapes such as elongated cuboids.[16-17]

Theoretical and computational research on chemical and optical properties of nanoclusters relies on examination of the electronic states and the corresponding wave functions computed from the density functional theory (DFT). As the properties and interrelations of the electronic states are closely related to the symmetries of the wave functions, it is beneficial to extract these symmetry representations.

For a long time, the convention in analyzing the symmetries of the wave functions in the Kohn-Sham (KS) DFT scheme of bare and ligand-protected metal clusters has been the projection of the wave functions to metal-core–centered spherical harmonics. The calculated weights of the various $Y_{lm}$ components in a given KS wave function are then used to characterize the "superatom character" of this particular KS wave function. The motivation lies in the superatom model, based on the spherically symmetric (on average) confining potential in which the electronic quantum states adapt similarly as the electron shells in a free atom.[18-23] In the ideal case of a perfect spherical symmetry, the allowed optical transitions can be evaluated directly from the angular momenta by using the dipole selection rule $\Delta l = \pm 1$. In practice, however, as the wave functions and consequently the electron density inherit the point group symmetry (if any) of the discrete atomic structure, this approach fails for shapes of the atomic structure that are far from spherical, or in the case where the atomic lattice interacts strongly with the delocalized electron gas of the metal, splitting and intermixing the angular momenta shells.

Generally, small bare metal clusters and also many ligand-protected metal clusters are expected to be electronically stabilized with an electron count (electronic "magic number") that closes the highest occupied angular momentum shell. This creates a distinct energy gap between the highest occupied and lowest unoccupied single-electron level (HOMO and LUMO, respectively). Larger clusters are expected to be stabilized by a favorable atomic packing of the metal, creating a series of atomic "magic numbers". Very recently, both stabilization mechanisms were demonstrated to be present simultaneously in cluster synthesis.[24] However, several known compositions and structures of ligand-protected gold and silver nanoclusters have strongly non-spherical core shapes and free-electron counts that do not match with expected electronic "magic numbers". Thus, deciphering the origin of the stabilization mechanisms of many known ligand-protected clusters creates continuing challenges to theory.

Attempts to generalize the "superatom" model[19] to take into account non-spherical shape and/or lattice effects are scarce. In 2017, we presented a scheme where the KS wave functions of the cuboidal-shape silver cluster $[Ag_{67}(SPhMe_2)_{32}(PPh_3)_8]^{3+}$ were projected onto the jellium wave functions of a 3D quantum box, which aided the assignment of symmetries based on box-quantization.[25] This method, however, was constrained to the cuboidal shape of the cluster core and required a reference calculation of the corresponding jellium box.

Here, we demonstrate the power of point-group based symmetry analysis of the electronic structure of both unprotected and ligand-protected metal nanoclusters. We assign point group symmetry representations for KS wave functions of two bare and seven ligand-protected Ag and Au nanoclusters: **(1)** $Ag_{55}^-$, **(2)** $Ag_{20}$, **(3)** $[Ag_{136}(TBBT)_{64}Cl_3]^-$ (TBBT = tert butyl benzene thiol), **(4)** $[Ag_{141}(SAd)_{40}Br_{12}]^+$ (SAd = adamantane thiol), **(5)** $Au_{70}S_{20}(PPh_3)_{12}$ (PPh$_3$ = tri phenyl phosphine), **(6)** $Au_{108}S_{24}(P(CH_3)_3)_{16}$, **(7)** $Au_{144}(SCH_3)_{60}$, **(8)** $[Au_{146}(p\text{-MBA})_{57}]^{3-}$ (p-MBA = para mercapto benzoic acid) and **(9)** $Au_{246}(SPhCH_3)_{80}$. We refer to these systems later either by the metal atom count or by the compound number. We show that the point group symmetry analysis brings out the symmetry characteristics of the frontier orbitals of these clusters in a superior way compared to the conventional spherical harmonics –based analysis for all symmetries that are lower than $I_h$. Furthermore, we demonstrate significant savings in CPU time when the symmetry information is used inside the linear-response calculation of the optical absorption spectrum of **1**.

## RESULTS AND DISCUSSION

The wave functions and eigenenergies for the KS states were solved using the real-space DFT code package GPAW[26] (see details in Methods). Experimental crystal structures were used directly for clusters **3**[10], **4**[11], **5**[27], **8**[15], and **9**.[12] For **6**, the PPh$_3$ ligand used in the experiment[28] was replaced by a simpler P(CH$_3$)$_3$, after which the ligand layer was optimized but the Au and S positions were kept fixed in the crystal structure. Cluster **7** is the theoretical model structure Au$_{144}$(S(CH$_3$)$_3$)$_{60}$ proposed by Lopez-Aceveo *et al.* in 2009 (ref. 9).

**Projection to point group symmetries.** The symmetry of a wave function is characterized via a set of overlap integrals

$$I_p = \int \psi^\dagger \hat{T}_p \psi \, \mathrm{d}\tau \qquad \text{Eq. (1)}$$

Where $\hat{T}_p$ is the specific symmetry operator for operation $p$, such as rotation around the main axis. The character tables for each point group are based on these integrals for perfectly symmetric objects, and they are given in the Supporting Information, Table S1. As metal nanoclusters very rarely possess perfectly symmetrical atomic structure, the integrals practically never give the exact symmetries as denoted by character tables.

However, because the rows of a character table constitute a set of linearly independent basis vectors, we write the symmetry vector of the wave function (the vector consisting of the overlap integrals appointed with different operations) as a linear combination of the rows. Solving the linear equations gives the symmetry of the wave function in terms of numerical weights for each symmetry representation. While solving these linear equations, the rows for degenerate symmetries are normalized so that operating with the unit operator $E$ on a normalized wave function gives 1, *i.e.*, in practice the row elements are divided by the degeneracy of the row. Due to the properties of the irreducible character table matrix, the sum of the linear coefficients equals to the first element of the overlap vector corresponding to the unit operation $E$ and thus always giving 1. Weights that are determined this way for the point group symmetries are then compared to the conventional way of projecting the KS wave functions to spherical harmonics (Y$_{lm}$ functions) as discussed in Ref. 19.

**Bare clusters**. We first compared the performance of the point group symmetry (PGS) analysis to Y$_{lm}$ analysis for two bare metal clusters Ag$_{55}^-$ and Au$_{20}$ (Figure 1). The projection to symmetry operators was done in a volume adding up atomic volumes of a radius of 3.0 Å from each atom. The Y$_{lm}$ projections were done in a spherical volume of 12 Å radius. The ground-state atomic structure of both clusters in gas phase has been determined previously. Based on comparison of photoelectron spectra and DFT calculations, Ag$_{55}^-$ was determined to have an icosahedral (I$_h$) structure.[29] In addition to I$_h$ symmetry, we studied Ag$_{55}^-$ also in two other closed-shell atomic configurations, namely in cubo-octahedral (O$_h$) and decahedral (D$_{5h}$) symmetries. For Au$_{20}$, we studied the tetrahedral T$_d$ structure that was first suggested for the Au$_{20}$ anion based on photoelectron spectroscopy data.[30] Later, it was also determined for the neutral Au$_{20}$ based on experimental-theoretical study of IR vibrations.[31]

The comparison of the PGS analysis to the Y$_{lm}$ analysis is shown in Figure 2. The free-electron count of Ag$_{55}$ cluster anion is 56, *i.e.*, it is two electrons shy from filling a magic-number electron shell at 58 electrons in a spherical system. In the perfectly spherical electron gas model (jellium), this corresponds to state fillings of 1S$^2$ 1P$^6$ 1D$^{10}$ 2S$^2$ 1F$^{14}$ 2P$^6$ 1G$^{16}$. As can be seen in the top panel of Figure 1a, there is a set of well-defined discrete states between the upper edge of the Ag(4d) band (at about -3 eV) and the Fermi energy, displaying the spherical symmetries S, F, P, and G in the energetic order. These states correspond to the above jellium notations 2S, 1F, 2P, and 1G. However, one sees that the I$_h$ symmetry splits the 1F and 1G shells very strongly. This was already noted in the early photoelectron spectroscopy study.[29] In the proper PGS I$_h$ analysis (lower panel of Figure 1a), the split shells are identified as 1F$^{14}$ → T$_{2u}$(6) + G$_u$(8) and 1G$^{16}$ → H$_g$(10) + G$_g$(6), where the electron numbers are shown in parenthesis in the symmetry notation. The decahedral cluster (Figure 2b) is seen to split almost all of the free-electron states very strongly, as revealed by the Y$_{lm}$ analysis. The D$_{5h}$ PGS analysis is successful in assigning the proper symmetry-dependent labels to these states, and the spherical symmetries are seen to split as follows: 1F$^{14}$ → A$_2$"(2) + E$_2$"(4) + E$_1$"(4) + E$_2$'(4) and 2P$^6$ → A$_2$"(2) + E$_1$"(4). The major highly degenerate peak of 1G$^{16}$ closest to the Fermi energy is seen to consist of D$_{5h}$-symmetric E$_2$', E$_1$", and E$_2$" states. The cubo-octahedral cluster is PGS-analyzed in O$_h$ symmetry and the analysis reveals the following splitting: 1F$^{14}$ → A$_{2u}$(2) + T$_{1g}$(6) + T$_{1u}$(6) and 1G$^{16}$ → E$_g$(4) + T$_{2g}$(6) + T$_{1g}$(6).

For the T$_d$ symmetric Au$_{20}$ cluster, there is only one identifiable free-electron state between the upper edge of the Au(5d) band (at about -1.5 eV) and the Fermi energy. The Y$_{lm}$ analysis yields the D-symmetry for the HOMO manifold (10 electrons), indicating that in this cluster the energy order in the spherical model between the 1D and 2S states is reversed. The T$_d$ PGS analysis further reveals that the highly degenerate HOMO manifold is split to E(4) and T$_2$(6).

When examining the d-band region in all systems, one sees a further interesting result. As expected, the Y$_{lm}$ fails in all cases in capturing the "global" symmetries of any d-band states, as they are very complicated linear combinations of atom-like d-orbitals. This is seen as the large gray areas in the PDOS in d-band regions in top panels of Figures 2a-d, which denote the electron density in the orbitals that cannot be described by the used spherical harmonics expansion (up to J-symmetry). However, we found out that the PGS analysis works very well for O$_h$ symmetric Ag$_{55}^-$ and T$_d$ symmetric Au$_{20}$ clusters, being able classify basically every state in the metal d-band to a given symmetry (see Figure 3 for O$_h$ Ag$_{55}^-$ and Figure S1 for Au$_{20}$). For I$_h$ and D$_{5h}$ Ag$_{55}^-$, the PGS analysis catches the symmetry of a large number of the d-band states (see Figures S2 and S3). This fact has an important consequence when we later discuss the use of generalized dipole selection rules for optical transitions and demonstrate how our PGS analysis can greatly reduce the computational cost in identifying the non-zero oscillator matrix elements in the linear-response calculation of optical absorption of O$_h$ Ag$_{55}^-$.

**Ligand-protected clusters.** The total structures and the structures of the metal cores of **3 – 8** are shown in Figure 4. The presence of ligands surrounding the metal core poses additional complications to the analysis of the electronic states in the metal core, since the electron density of a give KS state, while mostly residing in the core, may also spread out to ligands. Furthermore, the symmetry of the ligand layer may in some cases be lower than that of the metal core, as noted here for clusters **3** and **8**. This calls for judicious choices for selecting the volume(s) in which the overlaps with Y$_{lm}$ functions or with the point group operators are computed. The Y$_{lm}$ analysis needs a specification of a single sphere that reasonably contains the electron density in the metal core, and the chosen radii are given in Table 1 together with the point group symmetries. In the PGS analysis, we kept the same definition for the volume as in the case of bare clusters, *i.e.*, the overlaps to symmetry operators were calculated in a volume adding up atomic volumes of a radius of 3.0 Å from each core atom.

**Table 1.** Point group symmetries for the ligand-protected clusters studied in this work. For clusters **7** and **8**, we did the analysis of wave functions by using the symmetries shown in the parenthesis for the metal core.

|   |        | Core symm    | Ligand symm | R (Y$_{lm}$) (Å) |
|---|--------|--------------|-------------|------------------|
| 3 | Ag$_{136}$ | D$_{5h}$     | C$_2$       | 10               |
| 4 | Ag$_{141}$ | D$_5$        | D$_5$       | 10               |
| 5 | Au$_{70}$  | D$_{2d}$     | D$_{2d}$    | 9                |
| 6 | Au$_{108}$ | T$_d$        | T$_d$       | 9                |
| 7 | Au$_{144}$ | I (I$_h$)    | I (I$_h$)   | 11               |
| 8 | Au$_{146}$ | C$_2$ (C$_{2v}$) | C$_2$   | 9                |
| 9 | Au$_{246}$ | D$_5$        | D$_5$       | 12               |

As Figure 5b shows for cluster **4**, the calculated wave functions manifest the symmetry representations with great accuracy for a cluster with well-defined symmetry of the total structure, *i.e.*, where also the ligand layer possesses the symmetry (D$_5$) of the metal core. The assigned symmetries also show the degeneracy of the states correctly as the states labelled with E appear with higher degeneracy compared to the A symmetries. In contrast, **3** has a ligand layer that is of lower symmetry (C$_2$) than the 54-atom silver core (D$_{5h}$). Restricting the analysis to this smaller core gives rather clean symmetry states on both sides of the HOMO-LUMO energy gap (Figure 5a). The lower energy region (below –0.5 eV) can be ascribed to the ligand states with most of the electron density outside the analyzed volume and, consequently, the core symmetry analysis cannot assign any symmetry representations. For both **3** and **4**, the spherical angular momentum ($Y_{lm}$) analysis clearly indicates that the wave functions do not have spherical symmetry.

Clusters **5** and **6** are far from spherical and the $Y_{lm}$ analyses show no distinct features as expected, but the analyses based on the point group symmetry of the Au cores are very clean as shown in Figure 5c for **5** ($D_{2d}$) and Figure 5d for **6** ($T_d$). It is again notable that in both cases, the PGS analysis gives high weights also to the lower states that are within the Au(5d) band.

Figure 6a shows the results for the icosahedral cluster **7**. Analysis based on the $I_h$ group shows good performance in describing the symmetries of the states as it can attribute up to around 90% of the electron density to a single symmetry representation, while the corresponding ratio for $Y_{lm}$ analysis is around 60%. The deviations from perfect $I_h$ representations are most probably due to the imperfect icosahedral arrangement of the inner Au core and the slightly chiral arrangement of the 60 Au atoms at the core-ligand interface and the RS–Au–SR moieties in the ligand layer (in fact, the proper symmetry is the chiral icosahedral I) as noted already in 2009 when this structure model was proposed.[9] However, the analyses compare to each other very well, considering the $I_h$ point group based symmetries of the spherical harmonics.[32] The spherical symmetries of $Au_{144}$ around the Fermi level are S, D, H and I from the $Y_{lm}$ analysis, that correspond to S: $A_g$, D: $H_g$, H: $T_{1u}+T_{2u}+H_u$ and I: $A_g+T_{1g}+G_g+H_g$ in the $I_h$ representation. Our results are perfectly in line with this expected decomposition.

Regardless of the rather spherical shapes of clusters **8** and **9**, the $Y_{lm}$ projections fail in finding any proper character of the states (Figures 6b and 6c, respectively). Here again, PGS analysis based on the proper point group symmetry of their respective cores ($C_{2v}$ of **8** and $D_5$ of **9**) reveals clean symmetries of states in a wide energy range around the HOMO-LUMO gap.

**Point group specific selection rules for optical absorption.** The selection rules similar to the spherical rule $\Delta l = \pm 1$ can be devised for each character table. According to the Fermi's golden rule in quantum mechanics, the probability of an optical transition between two electronic states is proportional to the square of the transition dipole moment between the wave functions as

$$I \sim \left[ \int \psi_f^\dagger \hat{\mu}_k \psi_i \, \mathrm{d}\tau \right]^2 \qquad \text{Eq. (2)}$$

where $\hat{\mu}_k = -e\hat{k}$ is the dipole moment operator. The intensity goes trivially to zero if the integrand is antisymmetric, Thus, consideration of the symmetries $s_i$ and $s_f$ of the initial and final wave functions, respectively, is sufficient to determine if the transition is forbidden. Using the symmetry representations of a point group corresponding to the molecule in question, the integral in the equation (2) above becomes a sum of the products of the rows in the character table

$$I_{i \to f} \sim \left[ \sum s_f \circ \hat{\mu}_k \circ s_i \right]^2 \qquad \text{Eq. (3)}$$

where the sum is taken over the elements of the vector that results from the element-wise products, denoted by the symbol ∘. The dipole moment operator $\hat{\mu}_k$ only consists of the character table rows corresponding to the translational vectors. For example, in the $D_5$ point group, the translation along the main axis ($T_z$) has $A_2$ symmetry representation and the translations $T_x$ and $T_y$ have both $E_1$ representation. The allowed and forbidden transitions are determined directly by the symmetries: if the function inside the sum is antisymmetric, the sum over the values is 0, and the transition is always forbidden. Symmetric functions may lead to non-zero integral and to an allowed transition. This formulation also leads directly to the Laporte's rule,[33] stating that if the point group of a molecule has an inversion center, transitions are allowed only between states of which the other carries g (gerade) symmetry and the other has u (ungerade) symmetry. Transitions of types g → g and u → u are forbidden.

**Optical spectrum of $O_h$ $Ag_{55}^-$.** To generalize the selection rules for a point group over all directions, we used $\hat{\mu}_k = \hat{T}_x + \hat{T}_y + \hat{T}_z$ to tabulate the selection rules for the point group $O_h$ (Table S2), although in this point group the symmetry representation of each cartesian translation is the same, $T_{1u}$. In the table, the non-zero values from equation (3) are given as 1 (allowed) and the zero values are given as 0 (forbidden). These selection rules were then included in the linear-response time-dependent density functional theory (LR-TDDFT) calculation for the optical spectrum of the cubo-octahedral $Ag_{55}^-$ so that the optically forbidden transitions were removed from the calculation. Since the wave functions of $Ag_{55}^-$ carry very clean symmetries even in the Ag(4d) band (Figures 2c and 3), the forbidden transitions were straightforwardly defined: The states were assigned a single symmetry representation by their maximum symmetry weight, and a transition was excluded from the LR-TDDFT calculation if the selection rules denied the transition between these symmetry representations of the start and end states. The "symmetry-filtered" spectrum was practically identical to the one calculated without the symmetry-filtering, as seen in Figure 7. The run time of the symmetry-filtered calculation was reduced to 21% compared to the non-filtered calculation as a result from the fact that only 24 of the 100 inter-symmetry transitions are allowed in the $O_h$ point group. The very small differences of the spectra can be accounted for either by the numerical error due to the cartesian grid on which the wavefunctions are projected or by the small deviations of the atomic structure from the idealized point group symmetry.

## CONCLUSION

In this work, we have introduced an improved and generalized way to analyze electronic states of metal clusters that have nano-crystalline cores, *i.e.,* atomic arrangements with specific point group symmetries. For such systems, it is straightforward to calculate weights of each KS state (or "molecular orbital") projected to symmetry operators of the point group in question. We have shown that those electronic states of silver and gold clusters, both bare and ligand-protected, that reside mostly in the metal core, and close to the Fermi energy), are well classified to symmetry sub-groups by the PGS analysis. Furthermore, in many cases also the electron states in the metal d-band carry one major symmetry component with almost 100% weight. This has important consequences in the calculation of optical transitions via the linear response method, since it may allow significant savings in CPU time when symmetry-filtering of the states is done before calculating the elements in the oscillator matrix, as demonstrated here for the $O_h$ symmetric $Ag_{55}^-$ cluster.

We found that the $Au_{144}(SR)_{60}$ is the only cluster in this study for which the $Y_{lm}$ projection gives reasonably good weights, so it is rather close to shell filling orders according to the spherical electron gas (jellium) model. The high spherical symmetry of the states there can be attributed to the high symmetry of the atomic configuration: the icosahedral point group $I_h$ has 120 different symmetry operations, which can be regarded as a measure of how close the shape of the cluster is to a sphere. The second largest number of operations for point groups in this study is 48 for the octahedral $O_h$ group.

The strong splitting of jellium-type electron shells by the point group symmetry of the metal core in ligand-protected clusters can explain also the electronic stability of compounds where the electron counting rule[19] for "superatoms" yields a non-conventional electron number for spherical shell filling. The stabilization, that is, opening of the energy gap between the HOMO and LUMO states, is then a combined effect of the point group and the shape of the metal core. This effect can be surprisingly strong for clusters of size up to fairly large metal atom counts.

The rather large variety of systems analysed here, consisting of two different noble metals, several different ligands, as well as several sizes, shapes, and symmetries of the metal core, shows the generality of our approach, which has never before been applied to study the electronic structure of larger metal nanoclusters. Our analysis presented here can be applied to any nanoparticle with any shape that has a core of an identifiable point group symmetry, thus it can be viewed as the generalization of the superatom model introduced for spherical ligand-protected clusters ten years ago.[19]

## METHODS

**DFT and LR-TDDFT calculations.** The wave functions and eigenenergies for the KS states were solved using the real-space DFT code package GPAW.[26] The PBE (Perdew-Burke-Ernzerhof) functional[34] was used in all the calculations. The PAW setups for Ag and Au included relativistic effects. The wave functions were treated on a real-space grid with spacing of 0.20 Å. The systems were set in a computational cell with 5 Å of vacuum around the cluster. The structural optimization was deemed converged when the residual forces on atoms were below 0.05 eV/Å. The optical absorption spectrum of $Ag_{55}^-$ was calculated by using the LR-TDDFT module implemented in GPAW.[35] The PBE functional was used for the exchange-correlation kernel. The spacing of the real-space grid was 0.20 Å.

## ASSOCIATED CONTENT

The Supporting Information is available free of charge on the ACS Publications website at DOI:.. Content: Figures S1 – S3 and Tables S1, S2.

## AUTHOR INFORMATION

Corresponding Author: hannu.j.hakkinen@jyu.fi

Notes: The authors declare no competing interests.

## ACKNOWLEDGMENTS

This research was supported by the Academy of Finland (grant 294217 and H.H.'s Academy Professorship). The computational resources were provided by CSC – the Finnish IT Center for Science and by the Barcelona supercomputing center as part of the PRACE NANOMETALS project. S.K. thanks Vilho, Yrjö, and Kalle Väisälä Foundation for a PhD study grant.

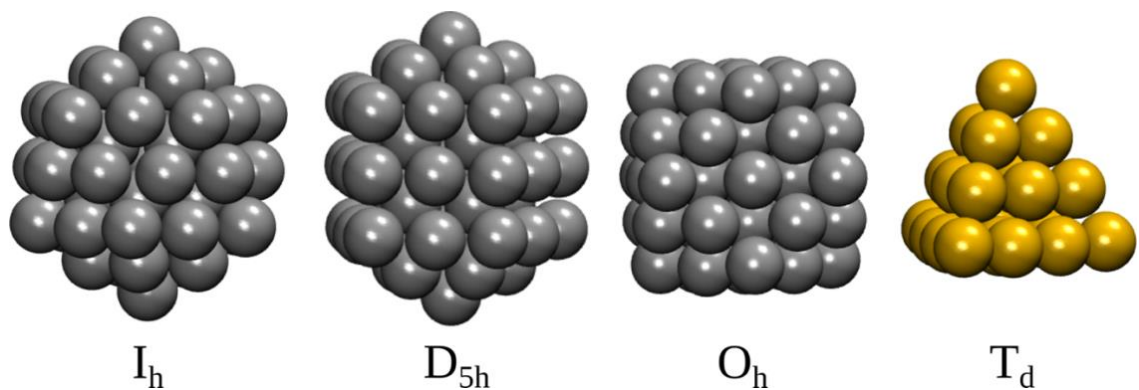

**Figure 1.** Bare metal clusters studied in this work. From left to right: icosahedral, decahedral, cubo-octahedral $Ag_{55}^-$, and tetrahedral $Au_{20}$, with shown point group symmetries.

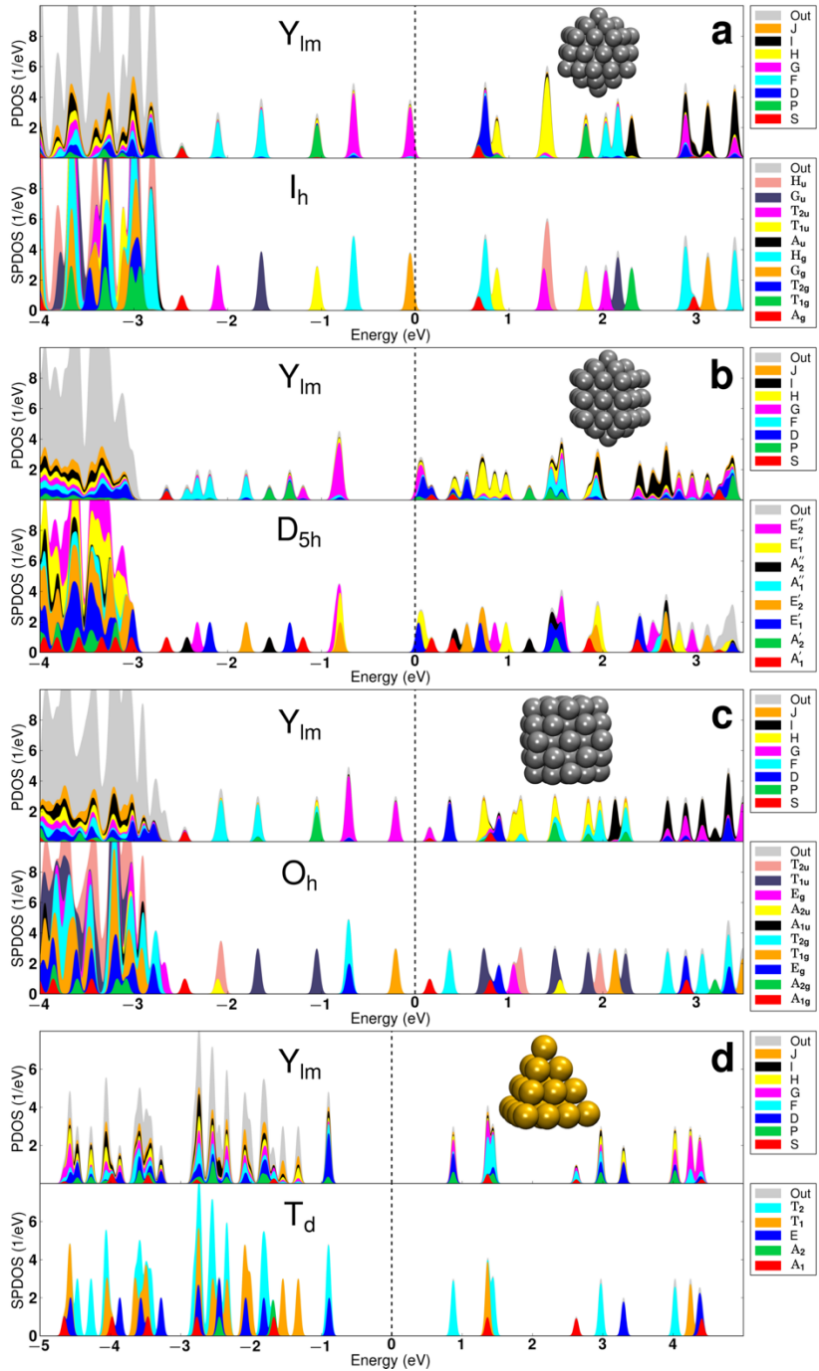

**Figure 2.** Analysis of KS states for clusters **1** and **2**. In each doublets of panels (a) – (d), the top panel shows the spherical harmonics ($Y_{lm}$) projected density of states (PDOS) and the bottom panel shows the PGS-analyzed DOS (SPDOS), with the point group symmetries shown in the panel. (a) icosahedral, (b) decahedral and (c) cubo-octahedral $Ag_{55}^-$, (d) tetrahedral $Au_{20}$. The DOS curves are obtained by broadening each discrete KS state with a 0.03 eV Gaussian. The Fermi energy is at zero. The band of Ag(4d) states starts around -3 eV in (a)-(c) and the band of Au(5d) states around -1.5 eV in (d). The grey area (denoted by label "out") shows electron densities that cannot be described by the projection to spherical harmonics (up to J symmetry) in the top panels or by the PGS analyses (within the chosen symmetry group) in the lower panels.

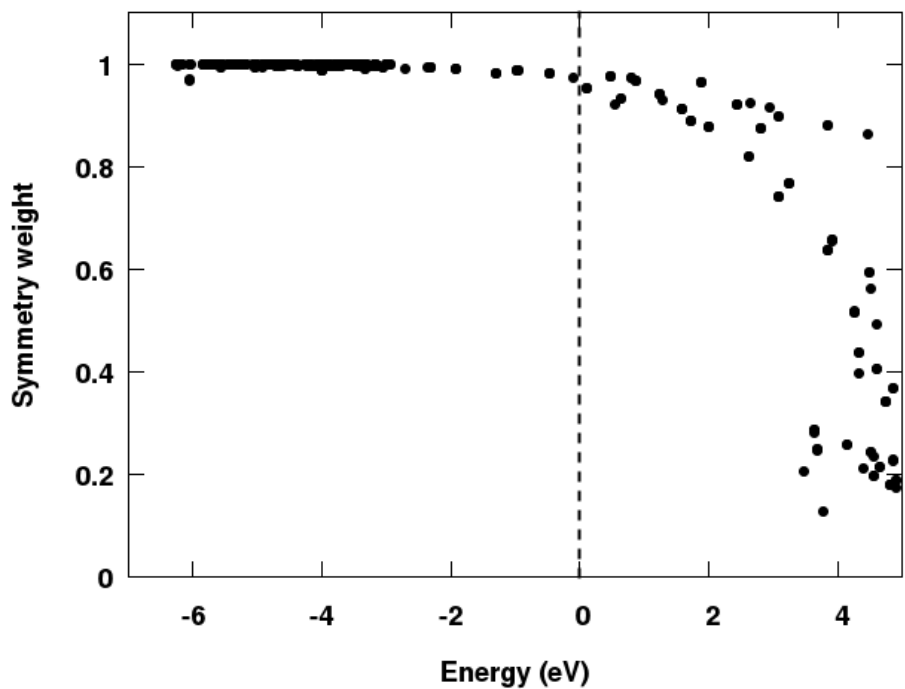

**Figure 3.** Maximum symmetry weights of each KS state of the cubo-octahedral $Ag_{55}^-$. Each dot represents one state. The Fermi energy is at zero. The Ag(4d)-band starts at around -3 eV.

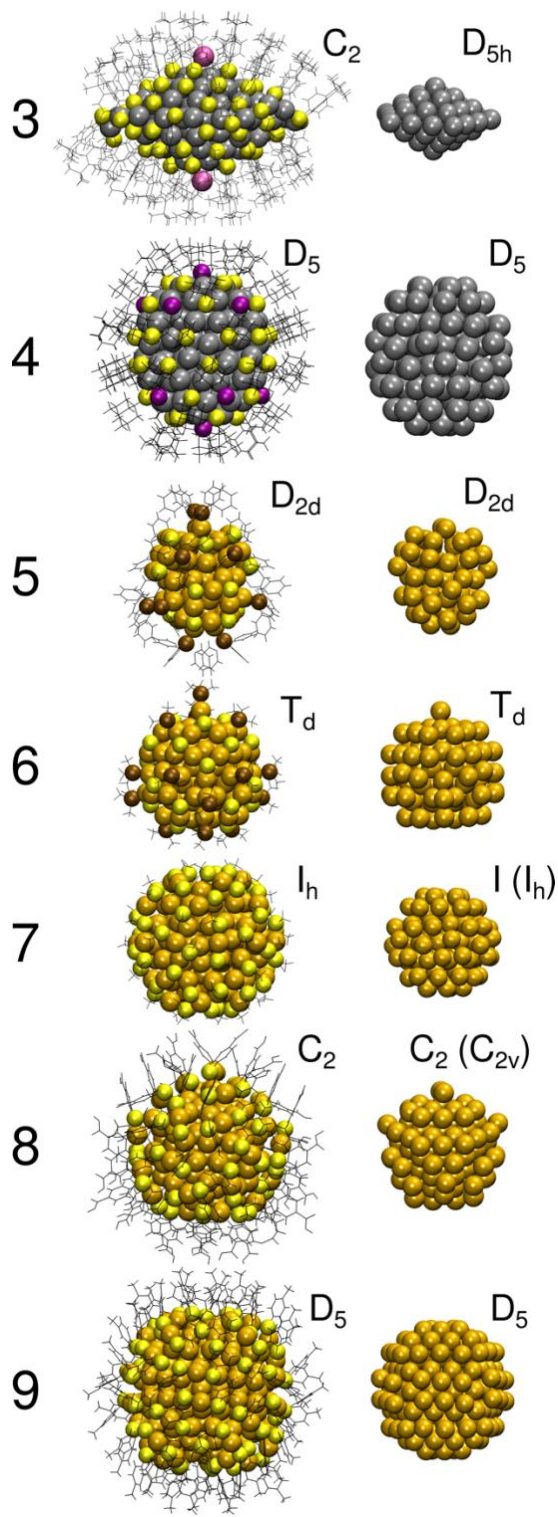

**Figure 4.** The clusters **3** to **9** (left to right, top row) and their metal cores with the point group assignments (bottom row). The main symmetry axis of each cluster lies in the vertical direction. See text for the chemical compositions. Ag: grey, Au: golden, S: yellow, P: brown, Cl: purple, Br: cyan. The ligand shells are indicated by the stick models.

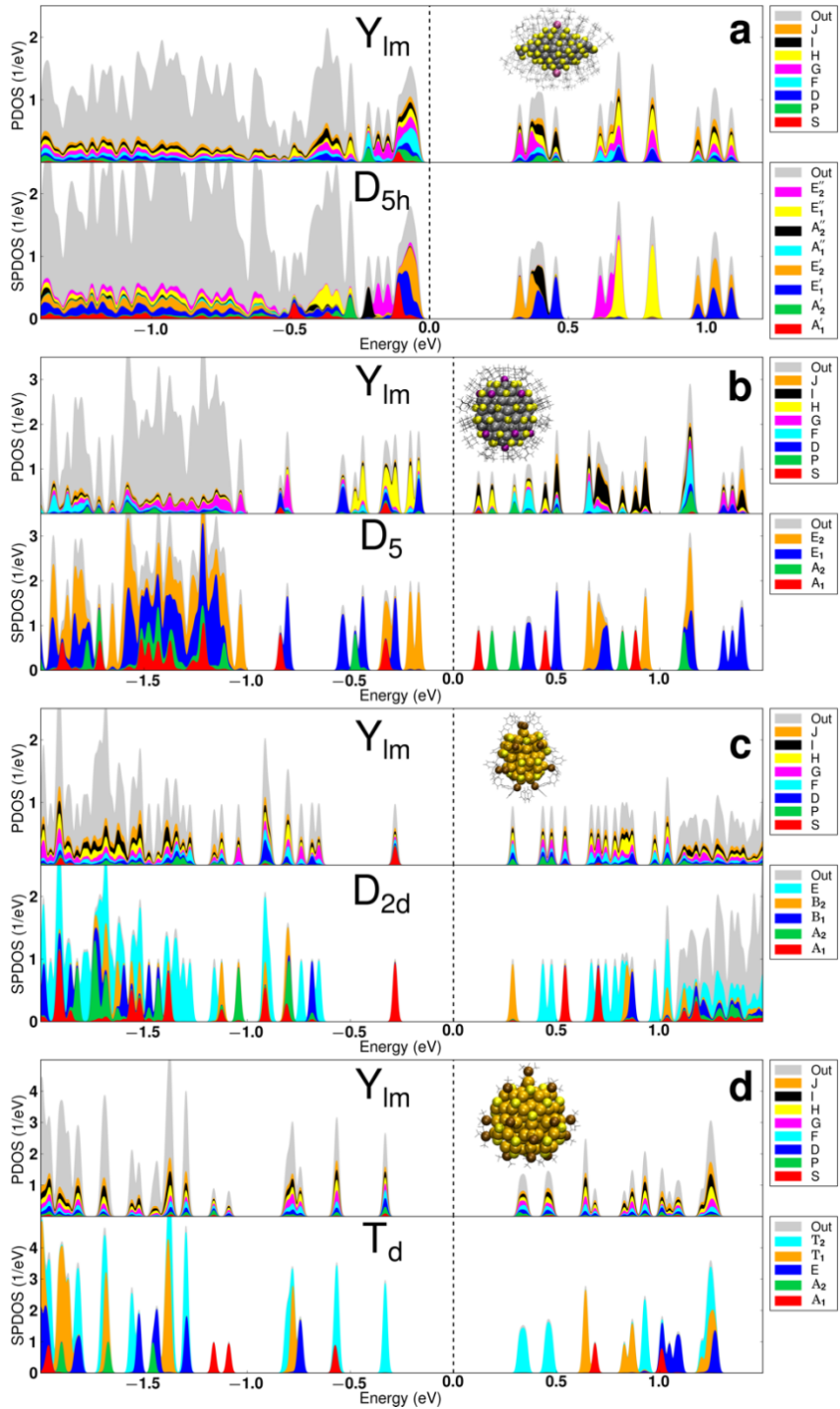

**Figure 5.** Analysis of the KS states for clusters (a) **3**, (b) **4**, (c) **5**, and (d) **6**. In each doublets of panels (a) – (d), the top panel shows the spherical harmonics ($Y_{lm}$) projected density of states (PDOS) and the bottom panel shows the PGS-analyzed DOS (SPDOS), with the point group symmetries shown in the panel. The grey area (denoted by label "out") shows electron densities that cannot be described by the projection to spherical harmonics (up to J symmetry) in the top panels or by the PGS analyses (within the chosen symmetry group) in the lower panels. The DOS curves are obtained by broadening each discrete KS state with a 0.01 eV Gaussian. The Fermi energy is at zero

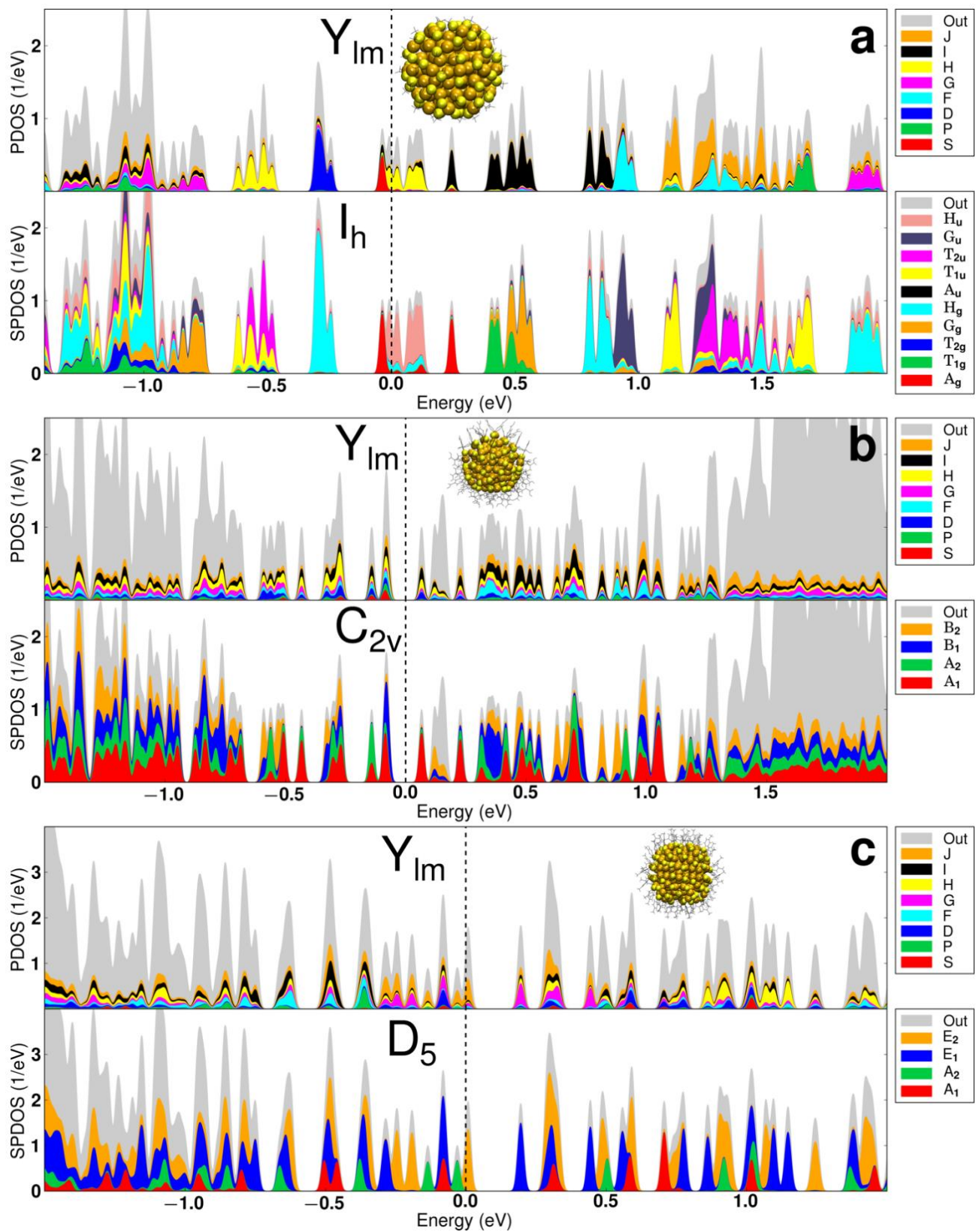

**Figure 6.** The same as Figure 5, but for clusters (a) **7**, (b) **8**, and (c) **9**.

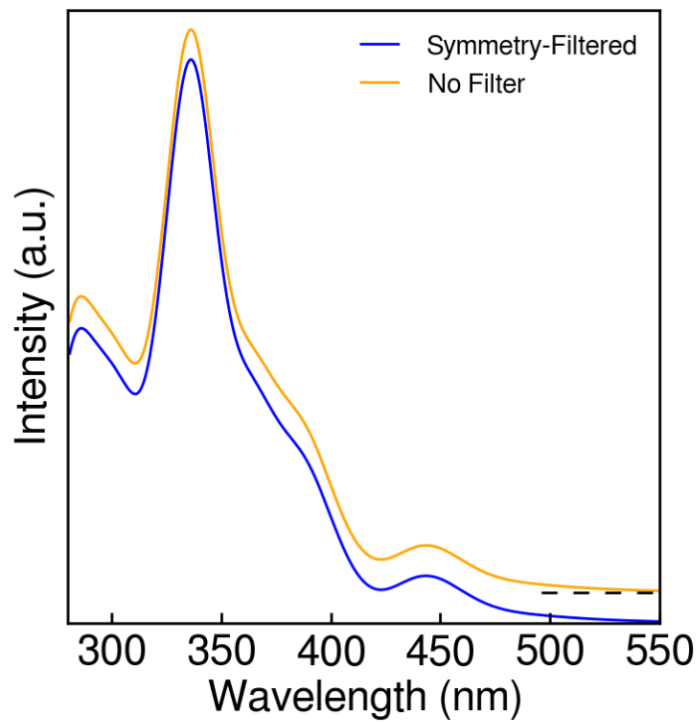

**Figure 7.** Computed optical spectra of the cubo-octahedral $Ag_{55}^-$ cluster by LR-TDDFT method. Brown curve: All transitions up to 4.5 eV included in the oscillator matrix, blue curve: only symmetry-filtered transition included. The brown curve is shifted up and decays to the dashed line.

**TOC GRAPHICS**

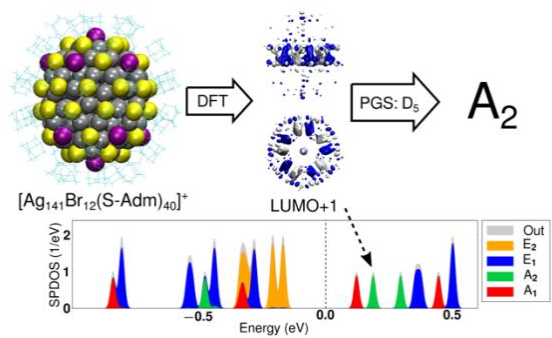

# Point Group Symmetry Analysis of the Electronic Structure of Bare and Protected Metal Nanocrystals

Sami Kaappa,[1] Sami Malola[1] and Hannu Häkkinen[1,2,*]

[1] Department of Physics, Nanoscience Center, University of Jyväskylä, FI-40014, Jyväskylä, Finland

[2] Department of Chemistry, Nanoscience Center, University of Jyväskylä, FI-400014, Jyväskylä, Finland

**Table S1.** Character tables for point group symmetries $D_5$, $D_{5h}$, $I_h$, $T_d$, $D_{2d}$, $C_{2v}$, and $O_h$.

$D_5$

|  | $E$ | $2\,C_5$ | $2\,(C_5)^3$ | $5\,C_2'$ |
|---|---|---|---|---|
| $A_1$ | 1 | 1 | 1 | 1 |
| $A_2$ | 1 | 1 | 1 | −1 |
| $E_1$ | 2 | $2\cos(2\pi/5)$ | $2\cos(4\pi/5)$ | 0 |
| $E_2$ | 2 | $2\cos(4\pi/5)$ | $2\cos(2\pi/5)$ | 0 |

$D_{5h}$

|  | $E$ | $2\,C_5$ | $2\,(C_5)^2$ | $5\,C_2'$ | $\sigma_h$ | $2\,S_5$ | $2\,(S_5)^3$ | $5\,\sigma_v$ |
|---|---|---|---|---|---|---|---|---|
| $A_1'$ | 1 | 1 | 1 | 1 | 1 | 1 | 1 | 1 |
| $A_2'$ | 1 | 1 | 1 | −1 | 1 | 1 | 1 | −1 |
| $E_1'$ | 2 | $2\cos(2\pi/5)$ | $2\cos(4\pi/5)$ | 0 | 2 | $2\cos(2\pi/5)$ | $2\cos(4\pi/5)$ | 0 |
| $E_2'$ | 2 | $2\cos(4\pi/5)$ | $2\cos(2\pi/5)$ | 0 | 2 | $2\cos(4\pi/5)$ | $2\cos(2\pi/5)$ | 0 |
| $A_1''$ | 1 | 1 | 1 | 1 | −1 | −1 | −1 | −1 |
| $A_2''$ | 1 | 1 | 1 | −1 | −1 | −1 | −1 | 1 |
| $E_1''$ | 2 | $2\cos(2\pi/5)$ | $2\cos(4\pi/5)$ | 0 | −2 | $-2\cos(2\pi/5)$ | $-2\cos(4\pi/5)$ | 0 |
| $E_2''$ | 2 | $2\cos(4\pi/5)$ | $2\cos(2\pi/5)$ | 0 | −2 | $-2\cos(4\pi/5)$ | $-2\cos(2\pi/5)$ | 0 |

$I_h$

| | E | 12 $C_5$ | 12 $(C_5)^2$ | 20 $C_3$ | 15 $C_2$ | i | 12 $S_{10}$ | 12 $(S_{10})^3$ | 20 $S_6$ | 15 $\sigma$ |
|---|---|---|---|---|---|---|---|---|---|---|
| $A_g$ | 1 | 1 | 1 | 1 | 1 | 1 | 1 | 1 | 1 | 1 |
| $T_{1g}$ | 3 | $-2\cos(4\pi/5)$ | $-2\cos(2\pi/5)$ | 0 | $-1$ | 3 | $-2\cos(2\pi/5)$ | $-2\cos(4\pi/5)$ | 0 | $-1$ |
| $T_{2g}$ | 3 | $-2\cos(2\pi/5)$ | $-2\cos(4\pi/5)$ | 0 | $-1$ | 3 | $-2\cos(4\pi/5)$ | $-2\cos(2\pi/5)$ | 0 | $-1$ |
| $G_g$ | 4 | $-1$ | $-1$ | 1 | 0 | 4 | $-1$ | $-1$ | 1 | 0 |
| $H_g$ | 5 | 0 | 0 | $-1$ | 1 | 5 | 0 | 0 | $-1$ | 1 |
| $A_u$ | 1 | 1 | 1 | 1 | 1 | $-1$ | $-1$ | $-1$ | $-1$ | $-1$ |
| $T_{1u}$ | 3 | $-2\cos(4\pi/5)$ | $-2\cos(2\pi/5)$ | 0 | $-1$ | $-3$ | $2\cos(2\pi/5)$ | $2\cos(4\pi/5)$ | 0 | 1 |
| $T_{2u}$ | 3 | $-2\cos(2\pi/5)$ | $-2\cos(4\pi/5)$ | 0 | $-1$ | $-3$ | $2\cos(4\pi/5)$ | $2\cos(2\pi/5)$ | 0 | 1 |
| $G_u$ | 4 | $-1$ | $-1$ | 1 | 0 | $-4$ | 1 | 1 | $-1$ | 0 |
| $H_u$ | 5 | 0 | 0 | $-1$ | 1 | $-5$ | 0 | 0 | 1 | $-1$ |

$T_d$

| | E | 8 $C_3$ | 3 $C_2$ | 6 $S_4$ | 6 $\sigma_d$ |
|---|---|---|---|---|---|
| $A_1$ | 1 | 1 | 1 | 1 | 1 |
| $A_2$ | 1 | 1 | 1 | $-1$ | $-1$ |
| E | 2 | $-1$ | 2 | 0 | 0 |
| $T_1$ | 3 | 0 | $-1$ | 1 | $-1$ |
| $T_2$ | 3 | 0 | $-1$ | $-1$ | 1 |

$D_{2d}$

| | E | 2 $S_4$ | $C_2$ | 2 $C_2'$ | 2 $\sigma_d$ |
|---|---|---|---|---|---|
| $A_1'$ | 1 | 1 | 1 | 1 | 1 |
| $A_2'$ | 1 | 1 | 1 | $-1$ | $-1$ |
| $B_1'$ | 1 | $-1$ | 1 | 1 | $-1$ |
| $B_2'$ | 1 | $-1$ | 1 | $-1$ | 1 |
| E | 2 | 0 | $-2$ | 0 | 0 |

$C_{2v}$

|       | E | $C_2$ | $\sigma_{v1}$ | $\sigma_{v2}$ |
|-------|---|-------|---------------|---------------|
| $A_1$ | 1 | 1     | 1             | 1             |
| $A_2$ | 1 | 1     | −1            | −1            |
| $B_1$ | 1 | −1    | 1             | −1            |
| $B_2$ | 1 | −1    | −1            | 1             |

$O_h$

|          | E | $8C_3$ | $6C_2$ | $6C_4$ | $3(C_4)^2$ | i  | $6S_4$ | $8S_6$ | $3\sigma_h$ | $6\sigma_d$ |
|----------|---|--------|--------|--------|------------|----|--------|--------|-------------|-------------|
| $A_{1g}$ | 1 | 1      | 1      | 1      | 1          | 1  | 1      | 1      | 1           | 1           |
| $A_{2g}$ | 1 | 1      | −1     | −1     | 1          | 1  | −1     | 1      | 1           | −1          |
| $E_g$    | 2 | −1     | 0      | 0      | 2          | 2  | 0      | −1     | 2           | 0           |
| $T_{1g}$ | 3 | 0      | −1     | 1      | −1         | 3  | 1      | 0      | −1          | −1          |
| $T_{2g}$ | 3 | 0      | 1      | −1     | −1         | 3  | −1     | 0      | −1          | 1           |
| $A_{1u}$ | 1 | 1      | 1      | 1      | 1          | −1 | −1     | −1     | −1          | −1          |
| $A_{2u}$ | 1 | 1      | −1     | −1     | 1          | −1 | 1      | −1     | −1          | 1           |
| $E_u$    | 2 | −1     | 0      | 0      | 2          | −2 | 0      | 1      | −2          | 0           |
| $T_{1u}$ | 3 | 0      | −1     | 1      | −1         | −3 | −1     | 0      | 1           | 1           |
| $T_{2u}$ | 3 | 0      | 1      | −1     | −1         | −3 | 1      | 0      | 1           | −1          |

**Table S2.** Optical selection rules for $O_h$ point group. Ones and zeros denote allowed and forbidden transitions, respectively.

|  | $A_{1g}$ | $A_{2g}$ | $E_g$ | $T_{1g}$ | $T_{2g}$ | $A_{1u}$ | $A_{2u}$ | $E_u$ | $T_{1u}$ | $T_{2u}$ |
|---|---|---|---|---|---|---|---|---|---|---|
| $A_{1g}$ | 0 | 0 | 0 | 0 | 0 | 0 | 0 | 0 | 1 | 0 |
| $A_{2g}$ | 0 | 0 | 0 | 0 | 0 | 0 | 0 | 0 | 0 | 1 |
| $E_g$   | 0 | 0 | 0 | 0 | 0 | 0 | 0 | 0 | 1 | 1 |
| $T_{1g}$ | 0 | 0 | 0 | 0 | 0 | 1 | 0 | 1 | 1 | 1 |
| $T_{2g}$ | 0 | 0 | 0 | 0 | 0 | 0 | 1 | 1 | 1 | 1 |
| $A_{1u}$ | 0 | 0 | 0 | 1 | 0 | 0 | 0 | 0 | 0 | 0 |
| $A_{2u}$ | 0 | 0 | 0 | 0 | 1 | 0 | 0 | 0 | 0 | 0 |
| $E_u$   | 0 | 0 | 0 | 1 | 1 | 0 | 0 | 0 | 0 | 0 |
| $T_{1u}$ | 1 | 0 | 1 | 1 | 1 | 0 | 0 | 0 | 0 | 0 |
| $T_{2u}$ | 0 | 1 | 1 | 1 | 1 | 0 | 0 | 0 | 0 | 0 |

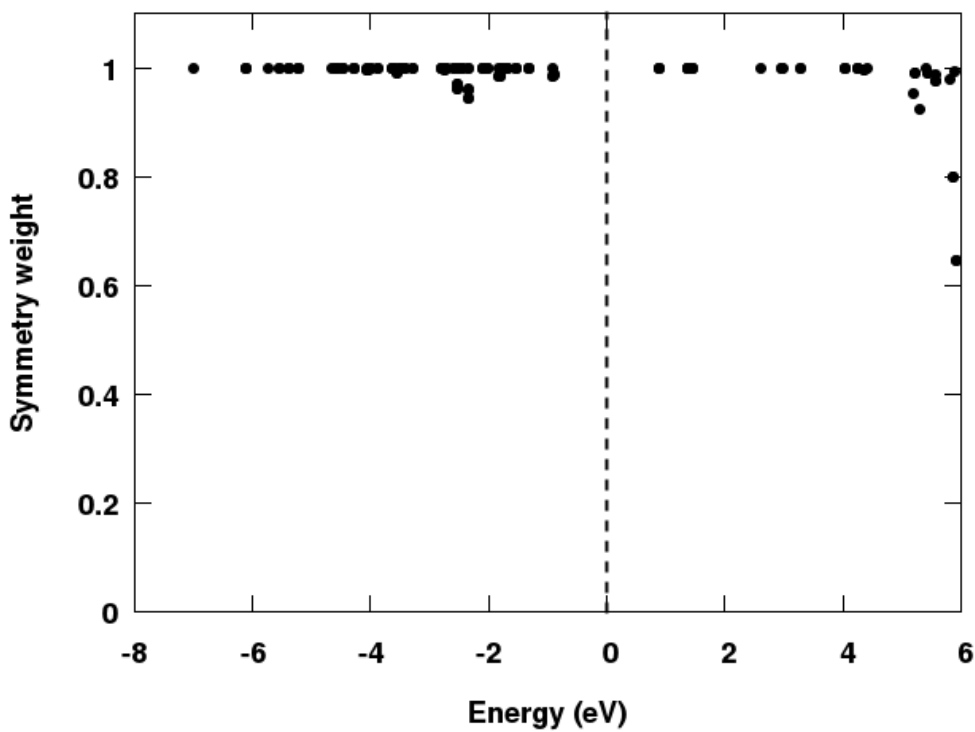

**Figure S1.** Maximum symmetry weights of each KS state of the $T_d$ $Au_{20}$. Each dot represents one state. The Fermi energy is at zero. The Au(5d)-band starts at around -1.5 eV.

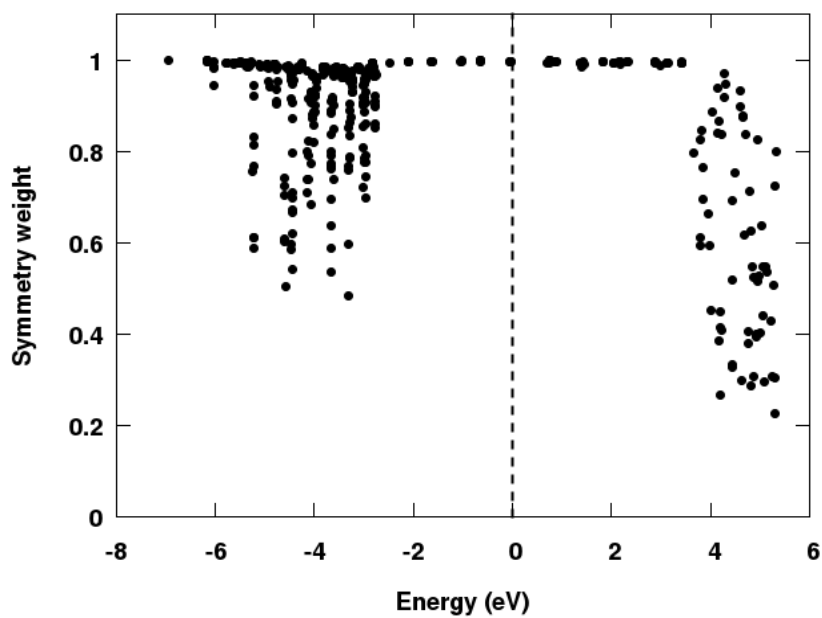

**Figure S2.** Maximum symmetry weights of each KS state of the $I_h$ $Ag_{55}^-$. Each dot represents one state. The Fermi energy is at zero. The Ag(4d)-band starts at around -3 eV.

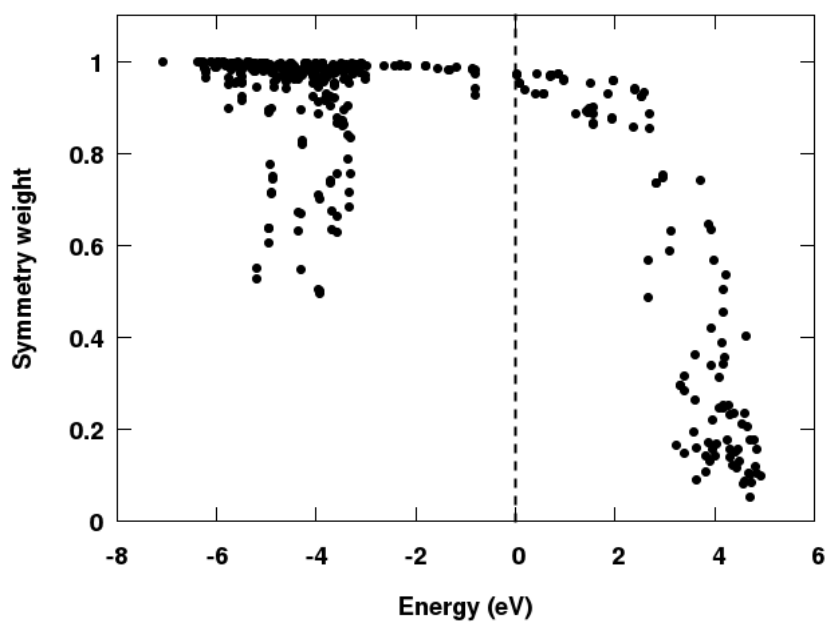

**Figure S3.** Maximum symmetry weights of each KS state of the $D_{5h}$ $Ag_{55}^-$. Each dot represents one state. The Fermi energy is at zero. The Ag(4d)-band starts at around -3 eV.